\documentclass[12pt]{article}

\usepackage{fullpage}
\usepackage{graphicx}
\usepackage{amsmath}
\usepackage{amssymb}
\usepackage{amsfonts}
\usepackage{amsthm}
\usepackage{lmodern}

\DeclareMathOperator*{\argmax}{arg\,max}

\def\ee{\mathbb{E}} 
\def\bSigma{\mbox{\boldmath $\Sigma$}}

\begin{document} 

\begin{center}

{\bf \large Stochastic simulators based optimization by Gaussian process metamodels - Application to maintenance investments planning issues} 
\vspace{0.5cm}

{\bf Short title: Metamodel-based optimization of stochastic simulators} 
\vspace{0.5cm}

{\bf Thomas BROWNE, Bertrand IOOSS, Lo\"ic LE GRATIET, J\'er\^ome LONCHAMPT, Emmanuel REMY}
\vspace{0.5cm}

EDF Lab Chatou, Industrial Risk Management Department
\vspace{0.5cm}

Corresponding author: Bertrand Iooss\\
EDF R\&D, 6 Quai Watier, F-78401 Chatou, France\\
Phone: +33130877969\\
Email: bertrand.iooss@edf.fr

\end{center}
 
\vspace{0.5cm}

\abstract{
This paper deals with the optimization of industrial asset management strategies, whose profitability is characterized by the Net Present Value (NPV) indicator which is assessed by a Monte Carlo simulator.
The developed method consists in building a metamodel of this stochastic simulator, allowing to get, for a given model input, the NPV probability distribution without running the simulator. 
The present work is concentrated on the emulation of the quantile function of the stochastic simulator by interpolating well chosen basis functions and metamodeling their coefficients (using the Gaussian process metamodel).
This quantile function metamodel is then used to treat a problem of strategy maintenance optimization (four systems installed on different plants), in order to optimize an NPV quantile. 
Using the Gaussian process framework, an adaptive design method (called QFEI) is defined by extending in our case the well known EGO algorithm. 
This allows to obtain an ``optimal'' solution using a small number of simulator runs. 
}
 
\vspace{0.5cm}


\section{Introduction}

The French company Electricit\'e de France (EDF) looks for assessing and optimizing its strategic investments decisions for its electricity power plants by using probabilistic and optimization methods of ``cost of maintenance strategies''. 
In order to quantify the technical and economic impact of a candidate maintenance strategy, some economic indicators are evaluated by Monte Carlo simulations using the VME software developed by EDF R\&D (Lonchamp and Fessart \cite{lonchamptfressart:2013}). 
The major output result of the Monte Carlo simulation process from VME is the probability distribution of the Net Present Value (NPV) associated with the maintenance strategy. 
From this distribution, some indicators, such as the mean NPV, the standard deviation of NPV or the regret investment probability ($\mathbb{P}(NPV < 0)$), can easily be derived.

Once these indicators have been obtained, one is interested in optimizing the strategy, for instance by determining the optimal investments dates leading to the highest mean NPV and the lowest regret investment probability. 
Due to the discrete nature of the events to be optimized, the optimization method is actually based on genetic algorithms.
However, during the optimization process, one of the main issues is the computational cost of the stochastic simulator to optimize (here VME), which leads to methods requiring a minimal number of simulator runs (Dellino and Meloni \cite{delmel15}).
Optimization algorithms require often several thousands of simulator runs and, in some cases, are no more applicable.

The solution investigated in this paper is a first attempt to break this computational cost by the way of using a metamodel instead of the simulator within the mathematical optimization algorithm. 
From a first set of simulator runs (called the learning sample and coming from a specific design of experiments), a metamodel consists in approximating the simulator outputs by a mathematical model (Fang et al. \cite{fanli06}). This metamodel can then be used to predict the simulator outputs for other input configurations, that can be served for instance for optimization issues (Forrester et al. \cite{forsob08}), safety assessment (de Rocquigny et al. \cite{derdev08}) or sensitivity analysis (Iooss and Lema\^{\i}tre \cite{ioolem15}).

Many metamodeling techniques are available in the computer experiments literature (Fang et al. \cite{fanli06}).
Formally, the function $G$ representing the computer model is defined as 
 \begin{equation}\label{eq:detmodel}
\begin{array}{rccl}
 G :& E & \rightarrow & \mathbb{R} \\
&x &\mapsto & G(x) 
\end{array}
\end{equation}
where $E \subset \mathbb{R}^d$ ($d\in \mathbb{N}\setminus \{0\}$) is the space of the input variables of the computer model.
Metamodeling consists in the construction of a statistical estimator $\widehat{G}$ from an initial sample of $G$ values corresponding to a learning set $\chi$ with $\chi \subset E$ and $\# \chi < \infty$ (with $\#$ the cardinal operator).
In this paper, we will use the Gaussian process (also called Kriging) metamodel (Sacks et al. \cite{sacwel89}) which has been shown relevant in many applicative studies (for example Marrel et al. \cite{marioo08}).
Moreover, this metamodel is the base ingredient of the efficient global Optimization (EGO) algorithm (Jones et al. \cite{jonsch98}), which allows to perform powerful optimization of CPU-time expensive deterministic computer code (Eq. (\ref{eq:detmodel})) as shown in Forrester et al. \cite{forsob08}.

However, all these conventional methods are not suitable in our applicative context because of the stochastic nature of the VME simulator: the output of interest is not a single scalar variable but a full probability density function (or a cumulative distribution function, or a quantile function).
The computer code $G$ is therefore stochastic:
\begin{equation}
\begin{array}{rccl}
G \, : \,& E\times\Omega & \rightarrow & \mathbb{R} \\
& (x,\omega) & \mapsto & G(x,\omega) 
\end{array}
\end{equation}
where $\Omega$ denotes the probability space. 
In the framework of robust design, robust optimization and sensitivity analysis, previous works with stochastic simulators consist mainly in approximating the mean and variance of the stochastic output (Bursztyn and Steinberg \cite{burste06}, Kleijnen \cite{kle15}, Ankenman et al. \cite{anknel10}, Marrel et al. \cite{marioo12}, Dellino and Meloni \cite{delmel15}).
Other specific characteristics of the distribution of the random variable $G(x,\cdot)$ (as a quantile) can also be modeled to obtain, in some cases, more efficient algorithms (Picheny et al. \cite{picgin13}).
In the following, $G(x)$, for any $x \in E$, denotes the output random variable $G(x,\cdot)$.

In this paper, as first proposed by Reich et al. \cite{reikal12} (who used a simple Gaussian mixture model), we are interested in a metamodel which predicts the full distribution of the random variable $G(x)$, $\forall x \in E$. 
We first focus on the output probability density functions of $G$ (\emph{i.e.} the densities of $G(x)$ for any input $x \in E$) as they are a natural way to model a random distribution.
Therefore we define the following deterministic function $f$:
\begin{equation}
\begin{array}{rccl}
f \, : \,& E & \rightarrow & \mathcal{F} \\
& x & \mapsto & f_x \quad \text{density of the r.v. $G(x)$}  \\
\end{array}
\end{equation}
with $\mathcal{F}$ the family of probability density functions:
 \begin{equation}  \mathcal{F} = \left\{ g \in L^1(\mathbb{R}), \; \text{positive, measurable}, \int_{ \mathbb{R} } g = 1 \right\} .  \end{equation} 
A first idea is to estimate the function $f$ over the input set $E$.
For a point $x \in E$, building $f_x$ with a kernel method requires a large number $N_{\mbox{\scriptsize MC}}$ of realizations of $G(x)$.
A recent work (Moutoussamy et al. \cite{mounan14}) has proposed kernel-regression based algorithms to build an estimator $\widehat{f}$ of the output densities, under the constraint that each call to $f$ is CPU-time expensive.
Their result stands as the starting point for the work presented in this paper.

The next section  presents the optimization industrial issues and our application case.
In the third section, we re-define the functions of interest as the output quantile functions of $G$ as they come with less constraints. 
We propose to use the Gaussian process metamodel and we develop an algorithm to emulate the quantile function instead of the probability density function. 
In the fourth section, this framework is used to develop a new quantile optimization algorithm, called Quantile Function Expected Improvement criterion and inspired from the EGO algorithm.
The normality assumptions set for the metamodel imply that the function to maximize, a given level of quantile, is also the realization of a Gaussian process.
In the following applicative section, this adaptive design method allows to obtain an ``optimal'' solution using a small number of VME simulator runs.
Finally, a conclusion synthesizes the main results of this paper.

\section{Maintenance investments planning issues and the VME tool}\label{sec:VME}

\subsection{Engineering asset management}

\begin{figure}[!ht]
\begin{center}
\includegraphics[width=0.8\textwidth]{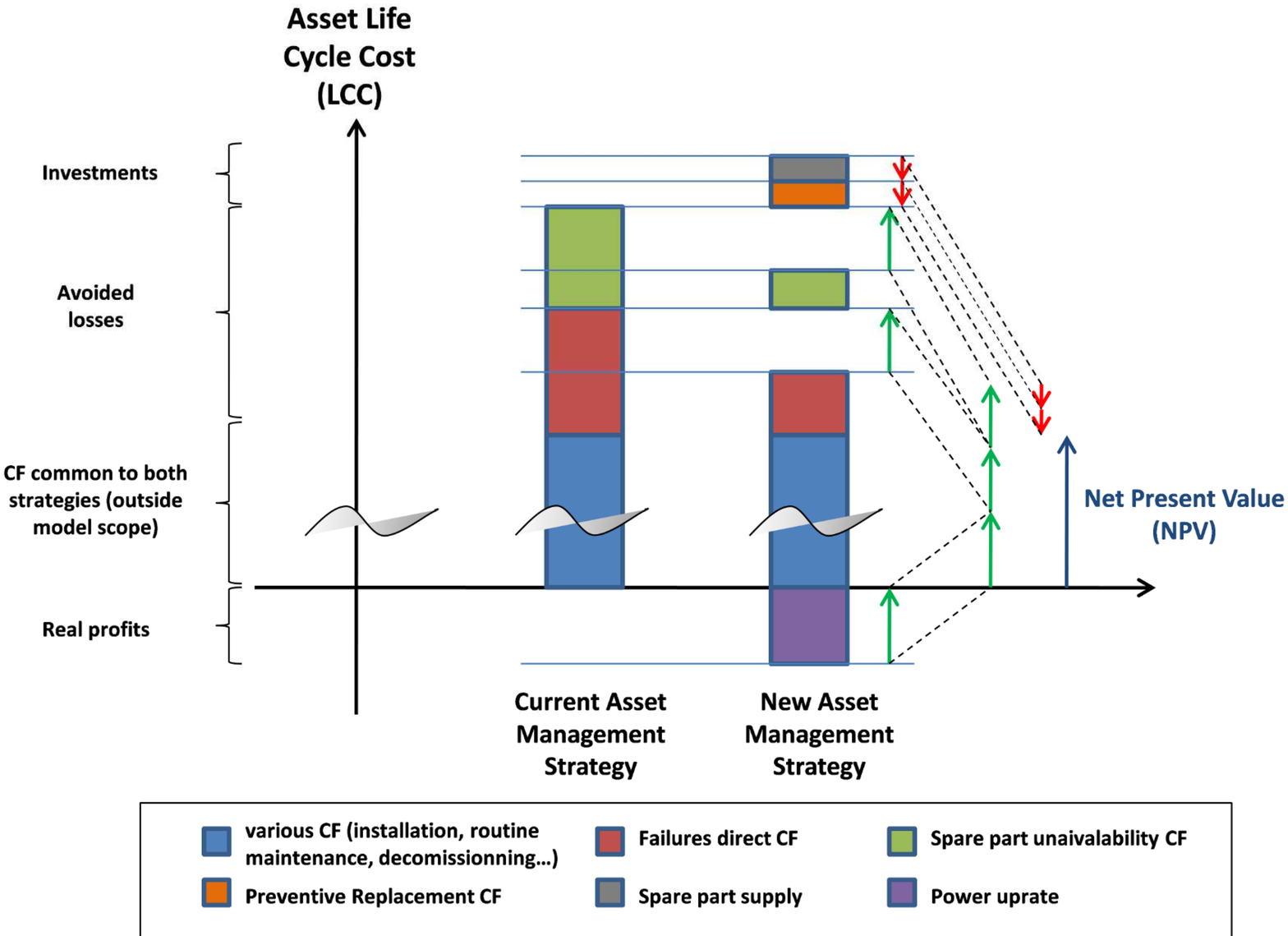}
\caption{Net Present Value of an investment in Engineering Asset Management.} 
\label{fig:NPV}
\end{center}
\end{figure}

Asset management processes, focused on realizing values from physical industrial assets, have been developed for years.
However, for the last one or two decades, these methods have been going from qualitative or semi-qualitative ones to quantitative management methods that are developed in the field of Engineering Assets Management (EAM). 
As a matter of fact, with budget issues becoming more and more constrained, the idea is not anymore to justify investments for the assets (maintenance, replacement, spare parts purchases \ldots) but to optimize the entire portfolio of investments made by a company or an organization.
The value of an asset may be captured in its Life Cycle Cost (LCC) that monetizes all the events that may impact an asset throughout its useful life in all its dimensions (reliability, maintenance, supply chain \ldots). 
The investments that are made for these assets (for example preventive replacements or spare part purchases) are then valorized by the variation of the LCC created by these changes in the asset management strategy. 
This variation is made of positive impacts (usually avoided losses due to avoided failures or stock shortages) and negative ones (investments costs). 
If the cash-flows used to calculate the LCC are discounted to take into account the time value of money, the value of an investment is then equivalent to a Net Present Value (NPV) as described in Figure \ref{fig:NPV}.

EAM is then about evaluating and optimizing the value of the asset management strategies in order to support investments decision making.

\subsection{VME case study}

\begin{figure}[!ht]
\begin{center}
\includegraphics[width=0.8\textwidth]{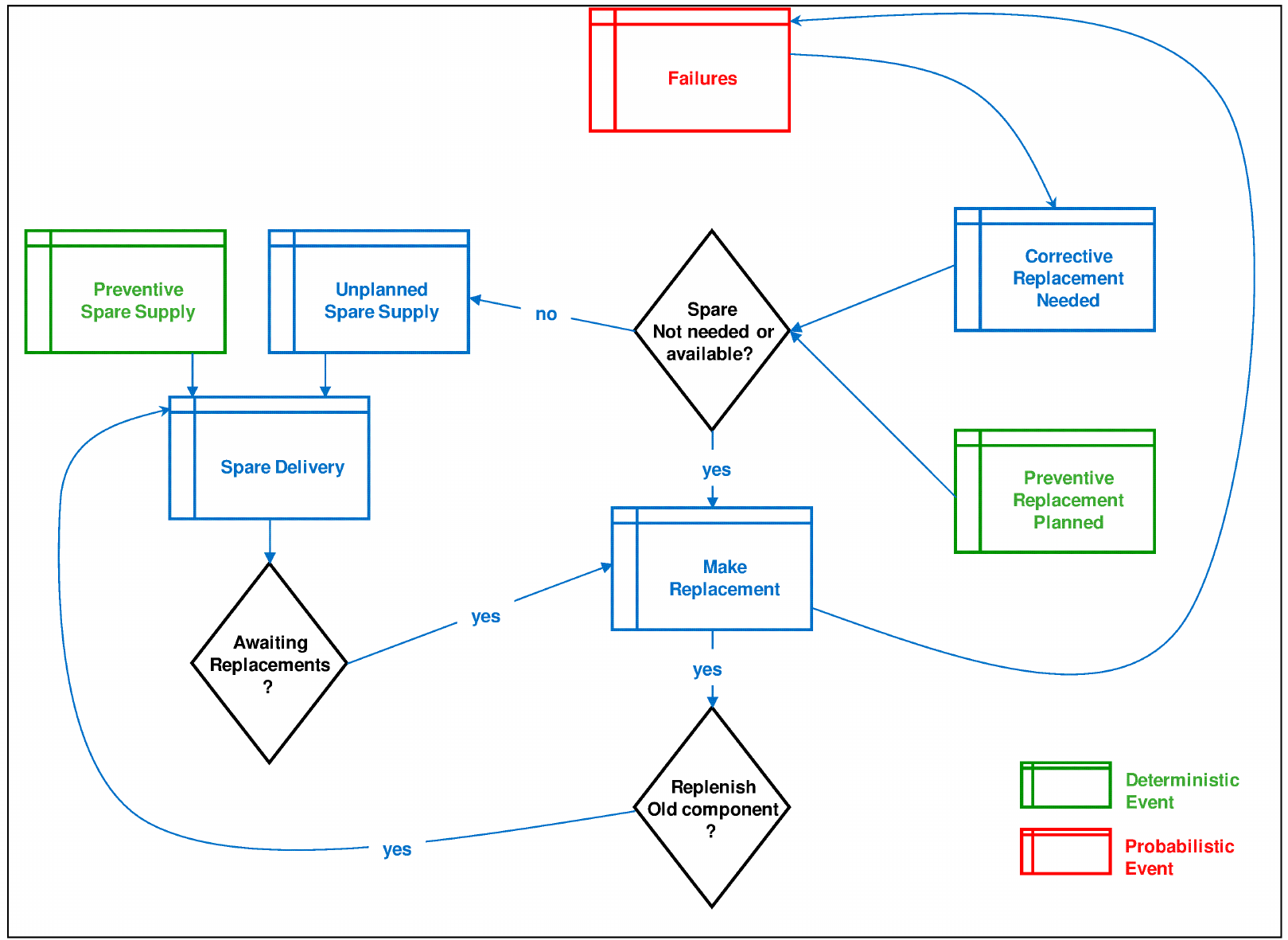}
\caption{Event model used in VME.} 
\label{fig:VME}
\end{center}
\end{figure}

As some of the cash-flows generated by the asset depend on stochastic events (failures dates depending on the probabilistic reliability), the NPV of an investment is also a stochastic variable that needs to be assessed. 
One of the tools developed by EDF to do so is a tool called VME that uses Monte Carlo simulation to evaluate various risk indicators needed to support decision making. 
The algorithm of VME consists in replicating the event model shown in Figure \ref{fig:VME} with randomized failure dates for both a reference strategy and the strategy to be evaluated in order to get a good approximation of the NPV probabilistic distribution.
Usually, the NPV takes some values in dollars or euros; for confidentiality reasons, no real cost is given in this paper and a fictive monetary unit is used.

One replication of the Monte-Carlo simulation consists in creating random dates for failures (according to reliability input data) and propagating them to the occurrence of the various events in the model (maintenance task, spare part purchase or further failure). 
The delay between these events may be deterministic (delay between the purchase of a spare and its delivery) or probabilistic (time to failure after maintenance), but both of them are defined by input parameters.  
Each event of the model generates cash flows (depending on input data such as the cost of a component, the value of one day of forced outage, \ldots) that pile up to constitute the Life Cycle Cost of the assets under one asset management strategy. 
The result of one replication is then the NPV that is the difference of the two correlated life-cycle costs of the current strategy and the assessed new strategy. 
This evaluation is replicated in order to obtain a good estimation of the probabilistic distribution of the NPV.

The test-case used in this paper is made of four systems installed on different plants, all these systems being identical (same reliability law) and using the same spare parts stock. 
The goal is to find the best replacement dates (in year) of these systems as it is wanted to prevent any failure event while replacements cannot be carried out too often.
When to purchase a new spare part also needs to be taken into account to secure the availability of the plants fleet (see Lonchamp and Fessart \cite{lonchamptfressart:2012} for more details).
This is given by the date (in year) of recovering a new system.
Therefore the whole strategy relies on these five dates which are taken as inputs for VME.
These dates can take different values (only discrete and integer values), which are described in Table \ref{tab:VMEinput}, and which have to be optimized.

\begin{table}
\center
\begin{tabular}{cccc}
\hline
Input & Name & min & max    \\
System replacement date for plant $1$ & $x_1$   & 41  & 50     \\
System replacement date for plant $2$ & $x_2$   & 41  & 50     \\
System replacement date for plant $3$ & $x_3$   & 41  & 50     \\
System replacement date for plant $4$ & $x_4$   & 41  & 50     \\
System recovering date & $x_5$   & 11  & 20     \\
\hline
\end{tabular}
\caption{Minimal and maximal values (in years) of the five inputs used in the VME model.}
\label{tab:VMEinput}
\end{table}

These five dates are the deterministic events displayed in Figure \ref{fig:VME} in green.
The random events, that make the NPV random, are the dates of failure of the plants.
They are illustrated in Figure \ref{fig:VME} in red. 
For a given set of five dates, VME computes a possible NPV based on a realization of the date of failure, randomly generated, regarding the different steps of the event model.
This input space of dates is denoted $E$ and regroups the possible years for replacements and the supply: 
	\begin{equation}  E=\left(\bigotimes _{i=1}^4 \left\lbrace 41,42,...,50 \right\rbrace  \right) \times  \left\lbrace 11,12,...,20 \right\rbrace .  \end{equation} 
$E$ is therefore a discrete set ($\# E = 10^5 $ ).
We have
	\begin{equation} \label{eq:densmodel}
\begin{array}{rccl}
G \, : \,& E\times\Omega & \rightarrow & \mathbb{R} \\
& (x,\omega)=(x_1,x_2,x_3,x_4,x_5,\omega) & \mapsto & \text{NPV}(x,\omega) , \\
\\
f \, : \,& E & \rightarrow & \mathcal{F} \\
& x=(x_1,x_2,x_3,x_4,x_5) & \mapsto & f_x \quad \text{(density of NPV}(x))  .\\
\end{array}
 \end{equation} 

\begin{figure}[!ht]
\begin{center}
\includegraphics[width=8cm,height=6cm]{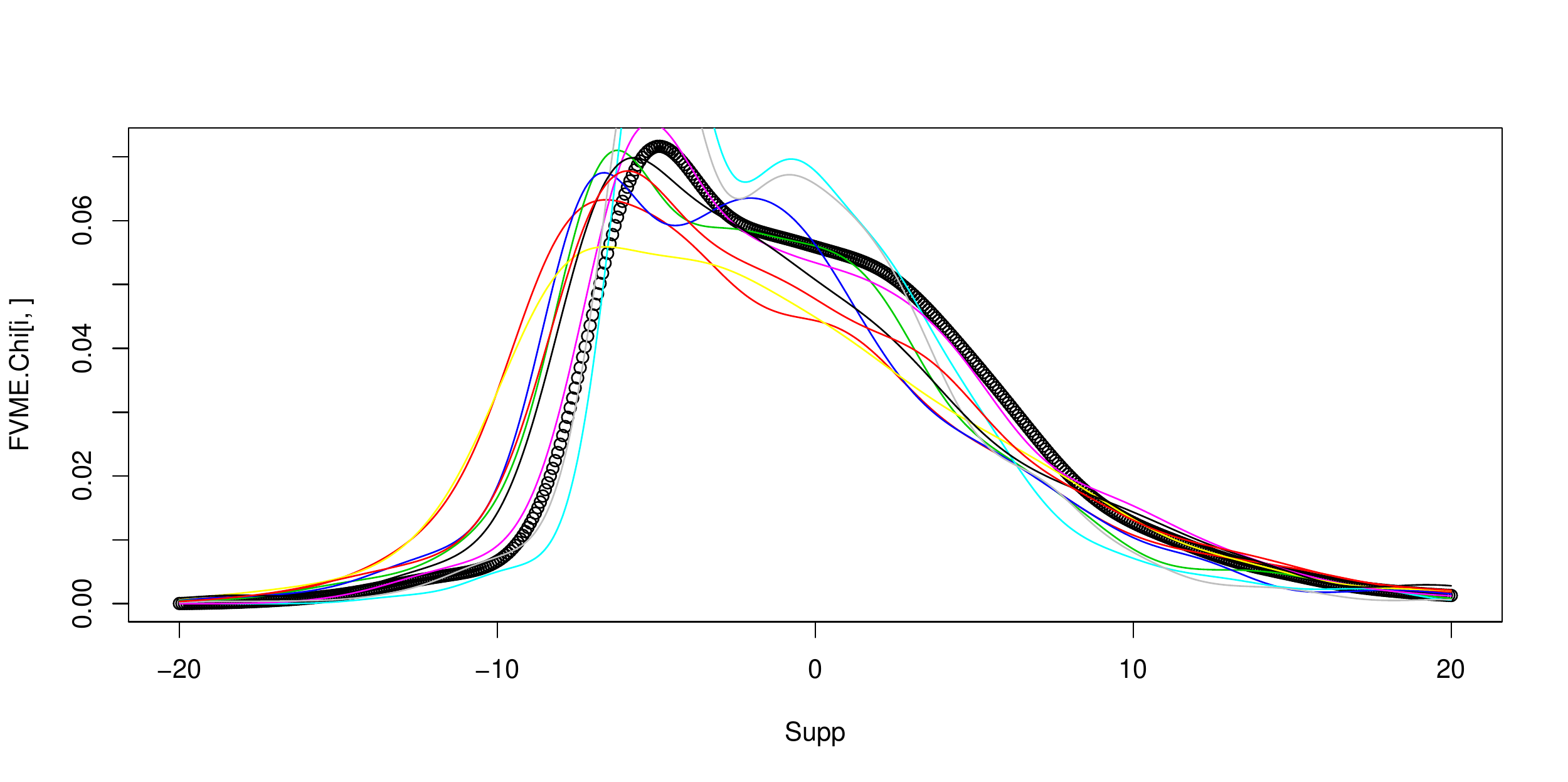}
\caption{$10$ output probability densities of the VME code (randomly sampled). In abscissa, the values of NPV are given using a fictive monetary unit.} 
\label{Opt_dens}
\end{center}
\end{figure}

Figure \ref{Opt_dens} provides examples of the output density of VME.
The $10$ input values inside $E$ have been randomly chosen.
It shows that there is a small variability between the curves.

The optimization process consists in finding the point of $E$ which gives the ``maximal NPV'' value. 
As NPV$(x)$ is a random variable, we have to summarize its distribution by a deterministic operator $H$, for example:
	\begin{equation}   
 H(g)=\ee(Z) \;,\; Z\sim g\;,\; \forall g \in \mathcal{F}
 \end{equation}   
or
	\begin{equation}   
 H(g)=q_Z(p)\;,\; Z\sim g\;,\; \forall g \in \mathcal{F}
 \end{equation}   
with $q_Z(p)$ the $p$-quantile ($0< p < 1$) of $g$.
Our VME-optimization problem turns then to the determination of
\begin{equation}  x^* := \argmax_{x \in E} H(f_x)  . \end{equation} 

However, several difficulties occur:
\begin{itemize}
\item VME is a CPU-time costly simulator and the size of the set $E$ is large. 
Computing $(f_x)_{x \in E}$, needing $N_{\mbox{\scriptsize MC}} \times \# E$ simulator calls (where $N_{\mbox{\scriptsize MC}}$ is worth several thousands), is therefore impossible.
Our solution is to restrict the VME calls to a learning set $\chi \subset E$ (with $\# \chi = 200$ in our application case), randomly chosen inside $E$.
We will then have $(f_x)_{x \in \chi}$.

\item
Our metamodel, that we will denote $\left(\hat{\hat{f_x}}\right)_{x  \in E} $, cannot be computed on $E$ due to its large size.
A new set $E$ is therefore defined, with $\#E = 2000$.
In this work, we limit our study to this restricted space (also denoted $E$ for simplicity) instead of the full space.
Other work will be performed in the future to extend our methodology to the study on the full space.
\end{itemize}

\section{Gaussian process  regression metamodel of a stochastic simulator}

\subsection{Basics on the Gaussian process regression model}

The Gaussian process regression model is introduced here in the framework of  deterministic scalar computer codes, following closely the explanations of Marrel et al. \cite{marioo08}.  
 The aim of a metamodel is to build a mathematical approximation of a computer code denoted by   $G(x) \in \mathbb{R}$, $x = (x_1,\ldots,x_d) \in E \subset \mathbb{R}^d$ from $n$ of its evaluations.
These  evaluations are performed from an experimental design set denoted in this work by  $\chi = ( x^{(1)},\ldots,x^{(n)})$ where $x^{(i)} \in E$, $i=1,\dots,n$. 
The simulations on  $\chi$  are denoted by $y_n = (y^{(1)},\ldots,y^{(n)})$ with $y^{(i)} = G(x^{(i)})$, $i=1,\dots,n$.

Gaussian process regression models (Sacks et al. \cite{sacwel89}), also called Kriging models (Stein \cite{ste99}), suppose that the simulator  response $G(x)$ is a realization of a Gaussian random process $Y$ of the following form:
\begin{equation}\label{Gpmodel}
 Y ( x ) = h ( x ) + Z ( x) ,
\end{equation}
where $h ( x ) $ is a deterministic function and $Z ( x)$ is a centered Gaussian process.

$h(x)$ gives the mean approximation of the computer code. In this work, $h(x)$ is  considered to be a one-degree polynomial model:
\begin{displaymath}  h ( x ) = \beta_0 + \sum_{j = 1}^d \beta_j x_j \;,  \end{displaymath} 
where $\beta = [ \beta_0, \beta_1,\ldots, \beta_d ]^t $ is a regression parameter vector.
The  stochastic part $Z(x)$ allows to model the non-linearities that  $h(x)$ does not take into account. Furthermore, the random process  $Z(x)$ is considered stationary with a covariance of the following form:
$$\mbox{Cov} ( Z ( x ), Z ( u ) ) = \sigma^2 K_{\theta} ( x - u ),$$
where $\sigma^2$ is the variance of $Z(x)$, $K_{\theta}$ is a correlation function and $\theta \in \mathbb{R}^d$ is a vector of hyperparameters.
This structure allows to provide interpolation and spatial correlation properties.
Several parametric families of correlation functions can be chosen (Stein \cite{ste99}).

Now let us suppose that we want to predict the output of $G(x)$ at a new point $x^{\Delta} = (x^{\Delta}_1,\ldots,x^{\Delta}_d) \in E$. The predictive distribution for $G(x^{\Delta})$ is obtained by conditioning $ Y ( x ^{\ast}) $ by the known values $y_n$ of $Y(x)$ on  $\chi$ (we use the notation $Y_n = Y(\chi) = (Y(x^{(1)}),\dots,Y(x^{(n)}))$. Classical results on Gaussian processes imply that this distribution is Gaussian:
\begin{equation}\label{pred_dist}
[  Y(x^{\Delta})|Y_n=y_n] = \mathcal{N}\left(\displaystyle \ee [  Y(x^{\Delta})|Y_n=y_n], \mbox{Var}[ Y(x^{\Delta})|Y_n] \right).
\end{equation}
Finally, the metamodel is given by the mean of the predictive distribution (also called kriging mean):
 \begin{equation}\label{eq_esperance}
 \displaystyle \ee [  Y(x^{\Delta})|Y_n=y_n] = h(x^{\Delta}) +  k(x^{\Delta})  ^t \bSigma_n^{-1} (Y_n - h(\chi)) 
 \end{equation}
 with
  \begin{displaymath}  
\begin{array}{lll}
k(x^{\Delta}) & = &[\mbox{Cov}(Y(x^{(1)}),Y(x^{\Delta})), \ldots, \mbox{Cov}(Y(x^{(n)}),Y(x^{\Delta}))  ] ^t  \\
& = & \sigma^2  [  K_{\theta} (x^{(1)},x^{\Delta}), \ldots, K_{\theta} (x^{(n)},x^{\Delta}) )  ]^t  
 \end{array} 
  \end{displaymath} 
  and 
\begin{displaymath}  \bSigma_n = \sigma^2 \left( K_{\theta} \left( x^{(i)} - x^{(j)} \right)  \right)_{i,j = 1\ldots n} \;. \end{displaymath} 
 Furthermore, the accuracy of the metamodel can be evaluated by the predictive mean squared error given by
 \begin{equation}\label{eq_variance}
 \displaystyle MSE(x^{\Delta}) = \mbox{Var}[ Y(x^{\Delta})|Y_n] = \sigma^2  -  k(x^{\Delta}) ^t  \bSigma_n^{-1} k(x^{\Delta}) \;
\end{equation}

The conditional mean \eqref{eq_esperance}  is  used as a predictor and its mean squared error (MSE)  \eqref{eq_variance} --  also called kriging variance -- is  a local indicator of the prediction accuracy. 
More generally, Gaussian process regression  model defines a Gaussian predictive distribution for the output variables at any arbitrary set of new points. 
This property can be used for uncertainty and sensitivity analysis (see for instance Le Gratiet et al. \cite{legcan14}). 

Regression and correlation parameters $\beta = (\beta_0, \beta_1,\dots,\beta_d)$, $\sigma^2$ and $\theta$ are generally  estimated with  a maximum  likelihood procedure (Fang et al. \cite{fanli06}). The maximum likelihood estimates of $\beta$ and $\sigma^2$ are given by the following closed form expressions:
\begin{displaymath}
\hat \beta = \left( h(\chi)^t \bSigma_n^{-1} h(\chi)\right)^{-1} h(\chi)^t \bSigma_n^{-1}y_n,  
\end{displaymath}
\begin{displaymath}
\hat \sigma^2 = \frac{\left(y_n - h(\chi)\hat \beta \right)^t \bSigma_n^{-1} \left(y_n - h(\chi)\hat \beta \right)}{n-(d+1)}.
\end{displaymath}
However, such expression does not exist for the estimate of $\theta$ and it has to be evaluated by solving the following problem (see for instance Marrel et al. \cite{marioo08}):
\begin{displaymath}
\hat \theta = \mathrm{argmin}_\theta \left[  \log(\mathrm{det}(\bSigma_n)) + n \log(\hat \sigma^2) \right].
\end{displaymath}
In practice, we have to rely on the estimators $\hat{h}$, $\hat{\beta}$ and $\hat{\theta}$. 
Therefore we introduce some extra variability in the computation of the Kriging mean and the variance.
For instance, this harms the normality distribution of $[  Y(x^{\Delta})|Y_n=y_n]$.
In \cite{zhu06} and \cite{zimmerman06}, the authors offer an accurate computation of the kriging variance based on a correction term.

\subsection{Emulation of the simulator quantile function}

\subsubsection{General principles}

\begin{figure}[!ht]
\begin{center}
\includegraphics[width=8cm,height=6cm]{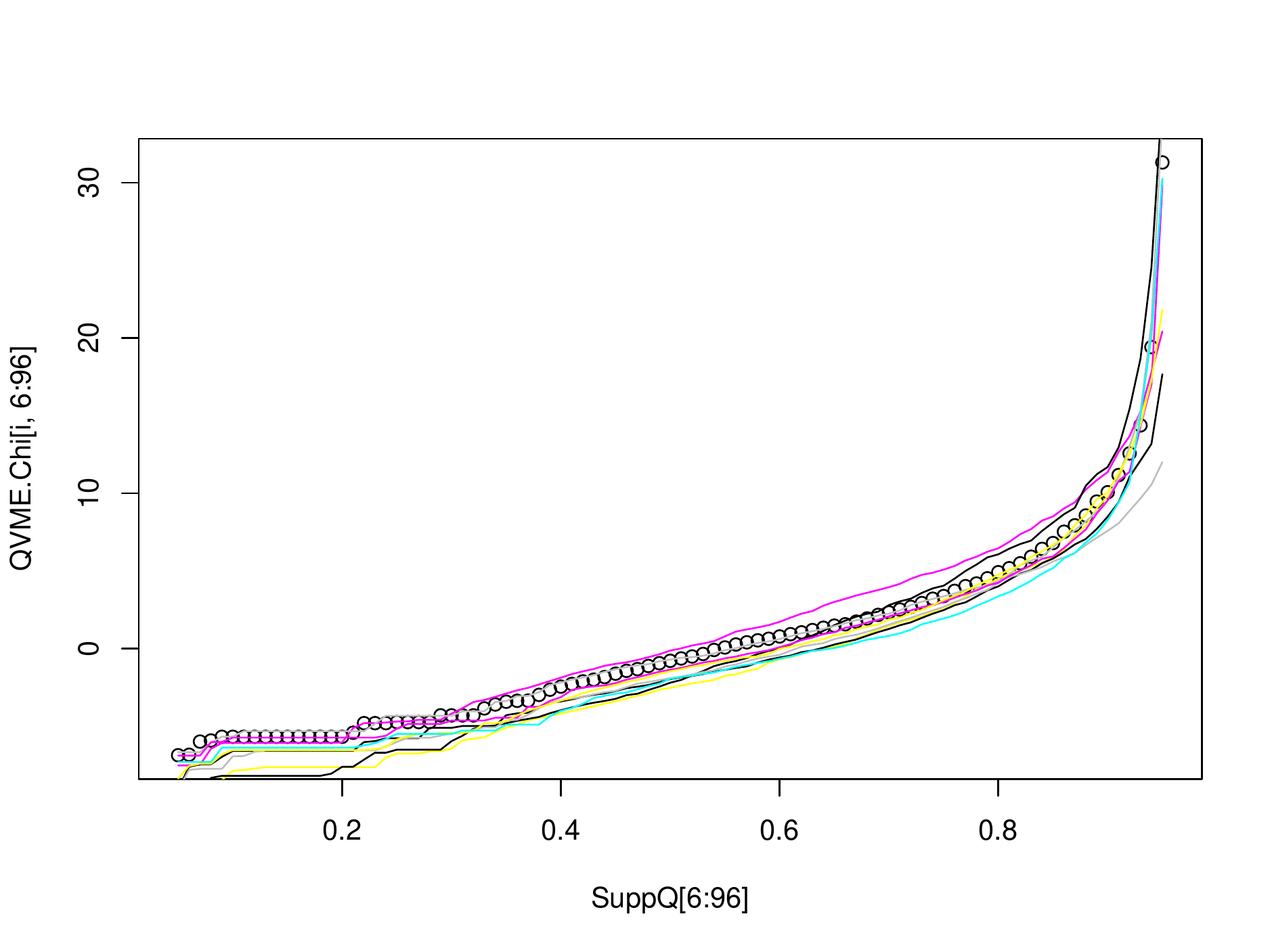}
\caption{The $10$ quantile functions $Q$ of the VME code associated with the $10$ VME pdf given in Figure \ref{Opt_dens}.} \label{plot5}
\end{center}
\end{figure}

In our VME-optimization problem, we are especially interested by several quantiles (for example at the order $1\%$, $5\%$, $50\%$, $95\%$, $99\%$) rather than statistical moments.
In Moutoussamy et al. \cite{mounan14} and Browne \cite{bro14}, quantile prediction with a density-based emulator has shown some deficiencies.
Mainly, both the positivity and integral constraints make it impossible to apply the following approach to output densities.
Therefore, instead of studying Eq. (\ref{eq:densmodel}), we turn our modeling problem to
\begin{equation} 
\begin{array}{rccl}
G \, : \,& E\times \Omega & \rightarrow & \mathbb{R} \\
& (x,\omega) = (x_1,x_2,x_3,x_4,x_5,\omega) & \mapsto & \text{NPV}(x,\omega) , \\
\\
Q \, : \,& E & \rightarrow & \mathcal{Q} \\
& x & \mapsto & Q_x \quad \text{quantile function of NPV$(x)$}  
\end{array}
 \end{equation} 
where $\mathcal{Q}$ is the space of increasing functions defined on $]0,1[$, with values in $[a,b]$ (which is the support of the NPV output). 
For $x \in E$, a quantile function is defined by
\begin{equation}  
\forall p \in ]0,1[, \;\; Q_x(p) = t \in [a,b] \quad \mbox{such that} \quad \int_a^t f_x(\varepsilon)d\varepsilon=p   .
\end{equation}  
For the same points as in Figure \ref{Opt_dens}, Figure \ref{plot5} shows the $10$ quantile function outputs $Q$ which present a rather low variability.

We consider the learning set $\chi$ ($n=\#\chi$) and $N_{\mbox{\scriptsize MC}} \times n$ $G$-simulator calls in order to obtain $\left( \tilde{Q}_x^{N_{\mbox{\tiny MC}}} \right)_{x \in \chi}$, the empirical quantile functions of $\left( \mbox{NPV}(x) \right)_{x \in \chi}$. 
In this work, we will use $N_{\mbox{\scriptsize MC}} = 10^4$, which is sufficiently large to obtain a precise estimator of $Q_x$ with $\tilde{Q}_x^{N_{\mbox{\tiny MC}}}$.
Therefore, we neglect this Monte Carlo error.
In the following, we simplify the notations by replacing $\tilde{Q}_x^{N_{\mbox{\tiny MC}}}$ by $Q_x$.

The approach we adopt is similar to the one used in metamodeling a functional output of a deterministic simulator (Bayarri et al., \cite{bayber07}, Marrel et. al. \cite{marioo11}).
The first step consists in finding a low-dimensional functional basis in order to reduce the output dimension by projection, while the second step consists in emulating the coefficients of the basis functions.
However, in our case, due to the nature of the functional outputs (quantile functions), some particularities will arise.

\subsubsection{Projection of $Q_x$ by the Modified Magic Points (MMP) algorithm}

Adapted from the Magic Points algorithm (Maday et al. \cite{madngu07}) for probability density functions, the MMP algorithm has been proposed in Moutoussamy et al. \cite{mounan14}.
It is a greedy algorithm that builds an interpolator (as a linear combination of basis functions) for a set of functions. Its principle is to iteratively pick a basis function in the learning sample output set and projecting the set of functions on the picked functions.
Its advantage over a more classical projection method (such as Fourier basis systems) is that it sequentially builds an optimal basis for this precise projection.

The first step of the algorithm  consists in  selecting in  the learning sample output set the functions which are the most correlated with the other ones. This function constitutes the first element of the functional basis. 
Then, at each step $j \in \{2, \ldots , k\}$ of the building procedure,  the element of the learning sample output set that maximizes the $ L^2$ distance  between itself and the interpolator --- using the previous $j-1$   basis functions --- is added to the functional basis.
The total number $k$ of functions is chosen with respect to a convergence criterion.
Mathematical details about this criterion are  not provided in the present paper.

In this paper, we apply the MMP algorithm on quantile functions.
Therefore, any quantile function  $\left( Q_x \right)_{x \in \chi}$ is expressed as follows:
\begin{displaymath} \hat{Q}_x = \sum_{j=1}^k \psi_j(x) R_j \;,\; \forall x \in \chi ,
 \end{displaymath} 
where $R=\left( R_1,...,R_k \right)$  are the   quantile   basis functions determined by MMP and $\psi=(\psi_1,\ldots,\psi_k)$ are the coefficients obtained by the projection of $Q_x$ on $R$.
In order to ensure the monotonic increase of $\hat{Q}_x$, we can restrict the coefficients  to the following constrained space:
\begin{displaymath}  
C = \left\lbrace \psi \in \mathbb{R}^k, \quad \psi_1,...,\psi_k \geq 0 \right\rbrace  .
\end{displaymath} 

However, this restriction is sufficient but not necessary.  That is why this constraint is not considered in Section \ref{optim_problem}. Indeed, it allows to preserve usefull  properties of Gaussian process metamodels (such as any linear combinations of Gaussian process metamodels is Gaussian)   for the optimization procedure. In practice, the monotonicity is verified afterwards. 

%
%
%

\subsubsection{Gaussian process metamodeling of the basis coefficients}\label{sec:metamodelcoef}
	
The estimation of the coefficients  $\psi(x) = (\psi_1(x), \ldots, \psi_k(x))$ ($x \in E$)  is performed with $k$ independent Gaussian process metamodels.
According to \eqref{pred_dist}, each $\psi_j(x)$, $j=1,\dots,k$, can be modeled by a Gaussian process of the following form:
	\begin{equation}\label{meta_psi}
 \psi_j(x) \sim \mathcal{N}\left( \hat{\psi}_j(x), MSE_j(x) \right) \;,\; \forall j \in \{1,...,k\} \;,\; \forall x \in E
  \end{equation}  
  where $\hat{\psi}_j(x)$ is the kriging mean \eqref{eq_esperance} and $MSE_j(x)$ the kriging variance \eqref{eq_variance} obtained from the $n$ observations ${\psi}_j(\chi)$.
  
Finally, the  following metamodel can be used for ${\hat{Q}}_x$:
\begin{equation}\label{eq:metamodel}
  \hat{\hat{Q}}_x = \sum_{j=1}^k \hat{\psi}_j(x)R_j  \;,\; \forall x \in E ,
	\end{equation} 

However, we have to ensure that $\hat{\psi}_j \in C$ $(j=1\ldots k$).
The logarithmic transformation can then be used:
\begin{displaymath} 
\begin{array}{rccl}
{\cal T}_1 \, : \,& C & \rightarrow & \mathbb{R}^k \\
& \psi & \mapsto & \left(\log(\psi_1+1),...,\log(\psi_k+1) \right) 
\end{array}
 \end{displaymath} 
 and its inverse transformation:
 \begin{displaymath} 
\begin{array}{rccl}
{\cal T}_2 \, : \,& \mathbb{R}^k & \rightarrow & C \\
& \phi & \mapsto & \left(\exp(\phi_1)-1,...,\exp(\phi_k)-1 \right). 
\end{array}
 \end{displaymath} 
Then supposing that $\phi(x) := {\cal T}_1(\psi(x))  \;,\; \forall x \in E$,  is a Gaussian process realization with $k$ independent margins, it can be estimated 
 by:
\begin{displaymath}
\hat{\phi}(x) = \ee[\phi(x)  \mid \phi(\chi) = {\cal T}_1(\psi(\chi))   ] \;,\; \forall x \in E ,
\end{displaymath}
 with $\psi(\chi)$ the learning sample output.
The following biased estimates of ${\psi}(x)$ can then be considered:
\begin{displaymath}
\hat{\psi}(x) = {\cal T}_2 (\hat{\phi}(x)) \;,\; \forall x \in E ,
\end{displaymath}
and (\ref{eq:metamodel}) can be applied as our metamodel predictor of the quantile function. However, the positivity constraint is not necessary and in our application the monotonicity is respected without considering it. Therefore, these transformations have not be used in our work.
%
%

\subsection{Numerical tests on a toy function}

\subsubsection{Presentation }

In order to test the efficiency of the estimator $\hat{\hat{Q}}$, we first substitute the industrial stochastic simulator VME by the following toy function $G$:
\begin{equation}
G(x) = \sin \left( x_1+U_1 \right) + \cos \left( x_2+U_2 \right) +x_3 \times U_3 , 
\label{eq:defG}
\end{equation}
with $x=(x_1,x_2,x_3) \in \left\lbrace 0.1; 0.2 ; ... ; 1 \right\rbrace^3$, $U_1 \sim \mathcal{N}(0,1)$, $U_2 \sim Exp(1)$ and $U_3 \sim \mathcal{U}([-0.5,0.5])$, which are all independent.  
$G(x)$ is a real random variable and we have $\# E=10^3$.
The goal is to estimate the output quantile functions:
$$
\begin{array}{rccl}
Q : \,& E & \rightarrow & \mathcal{Q} \\
& x & \mapsto & Q_x \quad \text{quantile function of $G(x)$} .
\end{array}
$$
As $G$ is a simple simulator and the input set $E$ has a low dimension ($d=3$), we can afford to compute $N_{\mbox{\scriptsize MC}}=10^4$ runs of $G(x)$ for each input $x \in E$.
Hence we can easily deduce all the output quantile functions $\left( Q_x \right)_{x \in E}$.

\begin{figure}[!ht]
\begin{center}
\includegraphics[width=8cm,height=6cm]{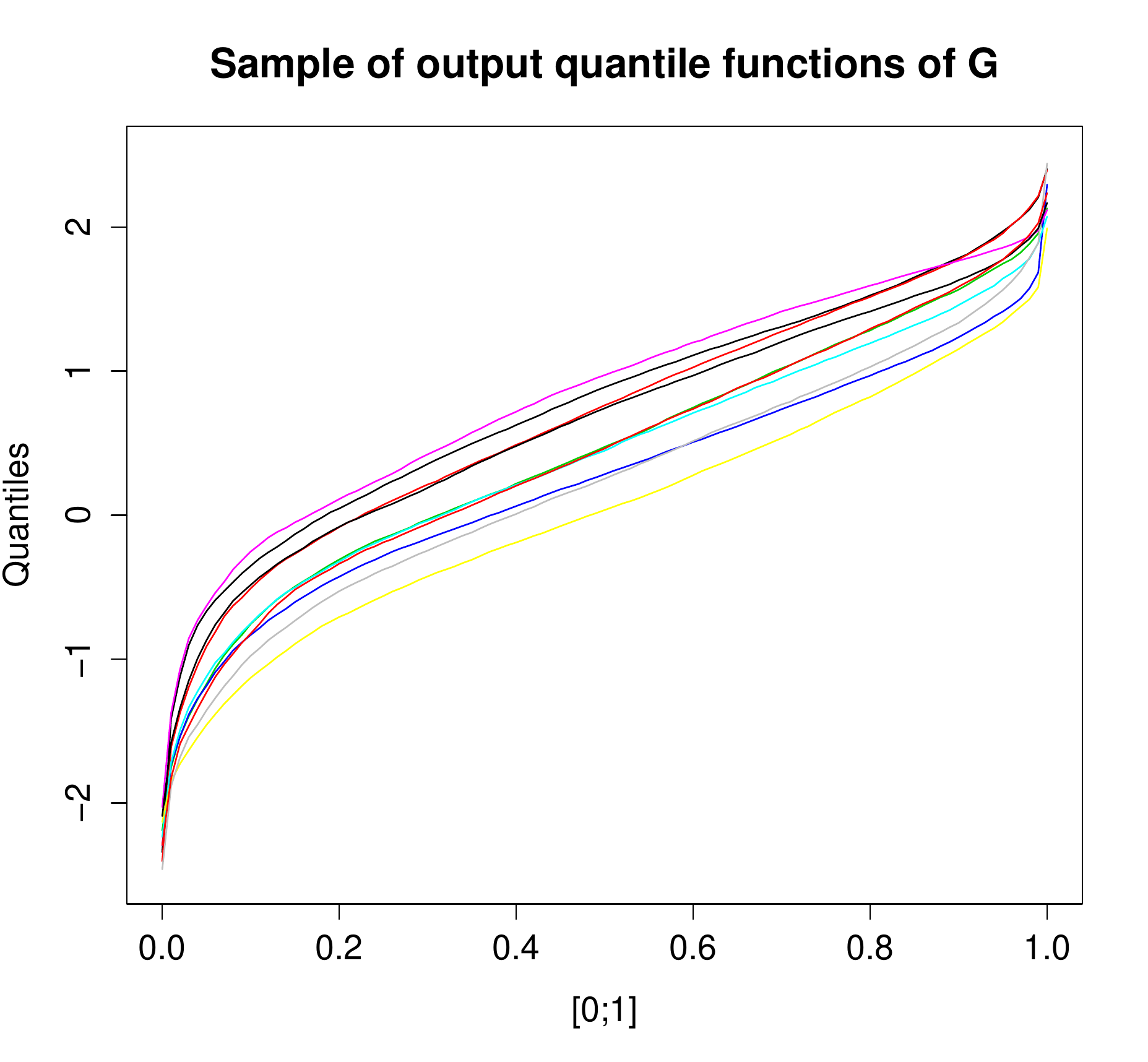}
\caption{$10$ ouput quantile functions of the simulator $G$ (randomly sampled).} 
\label{fig:Gquant_functions}
\end{center}
\end{figure}

We display in Figure \ref{fig:Gquant_functions} the output quantile functions $\left( Q_x \right)$ for 10 different $x \in E$.

\subsubsection{Applications to the MMP algorithm}

At present, we proceed at the first step for our estimator $\hat{\hat{Q}}$ by using the MMP algorithm.
We randomly pick the learning set $\chi \subset E$ such that $\# \chi = 150$.

As a result of the MMP algorithm, we get a basis of functions, $\left( R_1,...,R_k \right)$, extracted from the output quantile functions, $\left( Q_x \right)_{x \in E}$, as well as the MMP-projections of the output quantile functions on $\chi$:
$$
\forall x \in \chi\;, \quad \hat{Q}_x = \sum_{j=1}^k \psi_j(x) R_j.
$$
In the example, we set $k=4$.

Since the metamodel of the stochastic simulator is based on the estimation of the MMP-projection $\hat{Q}$, it is necessary to verify its relevance.
This is why we compute the following MMP error rate:
\begin{equation*}
err1 = \frac{1}{\# \chi} \: \sum_{x \in \chi} \:  \frac{\parallel  Q_x  - \hat{Q}_x  \parallel_{L^2} }{\parallel  Q_x  \parallel_{L^2} } = 0.09 \%.
\end{equation*}
This result shows that the MMP projection is very close to the real output quantile functions: if $\hat{\hat{Q}}$ is a good estimator of $\hat{Q}$, then it is a good estimator of $Q$.

It is important to recall that the basis $\left( R_1,...,R_k \right)$ depends on the choice of $\chi$, and so does $\hat{Q}$.
It probably approximates $Q$ better on $\chi$ than on the whole input set $E$.
Therefore it is natural to wonder whether the approximation $\hat{Q}$ is still relevant on $E$.
Since we could compute all the output quantile functions $\left( Q_x \right)_{x \in E}$, we could go further and compute
\begin{equation*}
err2 = \frac{1}{\# E} \: \sum_{x \in E} \:  \frac{\parallel  Q_x  - \hat{Q}_x  \parallel_{L^2} }{\parallel  Q_x  \parallel_{L^2} } = 0.13 \%.
\end{equation*}
From this result, we infer that even though the approximation is a little less correct on $E$, it is still efficient.

\subsubsection{Applications to the Gaussian process regression}\label{sec:applGP}

\begin{figure}[!ht]
\begin{center}
\includegraphics[width=8cm,height=6cm]{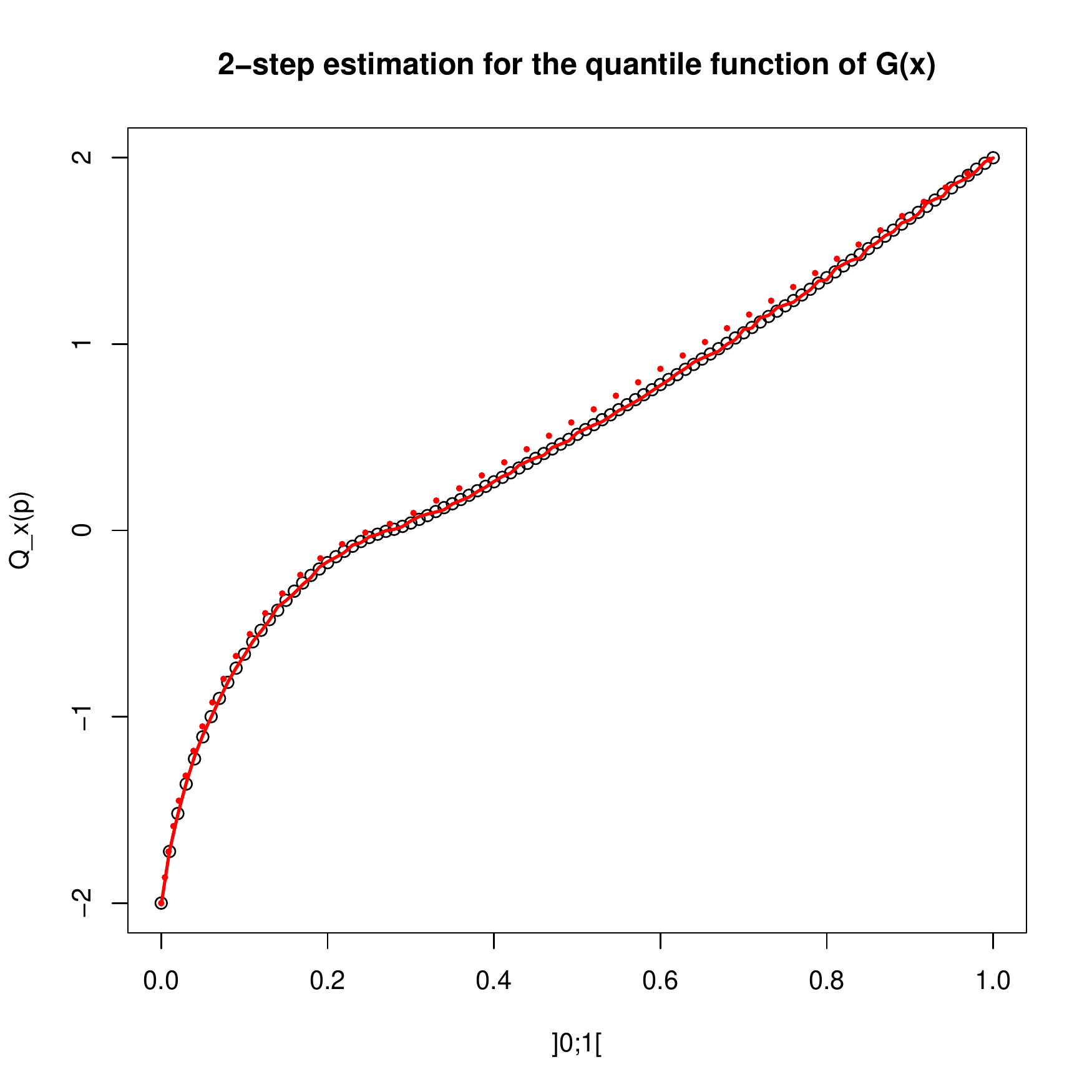}
\caption{For a point $x\in E$, $Q_x$ is in black points, $\hat{Q}_x$ in red line and $\hat{\hat{Q}}_x$ in red dotted line.} 
\label{fig:Gmetamodele}
\end{center}
\end{figure}

We now have the MMP projections $\left( \hat{Q}_x \right)_{x\in \chi}$, as well as the coefficients $\left( \psi_1(x),...,\psi_k(x) \right)_{x \in \chi}$.
The Gaussian process regression in this model consists in assuming that the $k$ coordinates $\left( \psi_1(x),...,\psi_k(x) \right)_{x \in E}$ are the realizations of $k$ independent Gaussian processes whose values on $\chi$ are already known.
From Eq. (\ref{eq:metamodel}), we have the expression of the metamodel:
$$
\forall x \in E\;, \quad \hat{\hat{Q}}_x = \sum_{j=1}^k \hat{\psi}_j(x) R_j .
$$
We verify the efficiency of the metamodel by computing the following error:
\begin{equation*}
err3 = \frac{1}{\# E} \: \sum_{x \in E} \:  \frac{\parallel  Q_x  - \hat{\hat{Q}}_x  \parallel_{L^2} }{\parallel  Q_x  \parallel_{L^2} } = 1.42 \%.
\end{equation*}
We conclude from this result, that the metamodel provides a good approximation for the output quantile functions of the simulator $G$ as runs were performed only on $\chi$.
We display in Figure \ref{fig:Gmetamodele} the output quantile function $Q_x$ for a point $x \in E$, with its successive approximations $\hat{Q}_x$ and $\hat{\hat{Q}}_x$.

Figure \ref{fig:Gmetamodele} confirms that, for this point $x\in E$ randomly picked, the MMP projection is really close to the initial quantile function.
As for $\hat{Q}_x$, the error is larger but the whole distribution of $G(x)$  is well estimated.

\section{Application to an optimization problem} \label{optim_problem}

\subsection{Direct optimization on the metamodel}

A first solution would be to apply our quantile function metamodel with a quantile-based objective function:
\begin{equation}   
\begin{array}{rccl}
H  \, : \,& \mathcal{Q} & \rightarrow & \mathbb{R} \\
& q & \mapsto & q(p) \\
\end{array}  
 \end{equation}   
with $p\in ]0,1[$.
We look for:
\begin{equation}  x^* := \argmax_{x \in E} Q_x(p) \end{equation} 
but have only access to:
\begin{equation}  \tilde{x}^* := \argmax_{x \in E} \hat{\hat{Q}}_x(p) . \end{equation} 
We also study the relative error of $H(\hat{\hat{Q}})$ on $E$ by computing:
\begin{equation}  
err = \frac{1}{\max_{x \in E} \left( Q_x(p) \right) -\min_{x \in E} \left( Q_x(p) \right) } \times \left( \sum_{x \in E} \mid Q_x(p) - \hat{\hat{Q}}_x(p) \mid \right) .
\end{equation} 

As an example for $p=0.5$ (median estimation), with the toy-function $G$ introduced in (\ref{eq:defG}), we find
\begin{equation*}   
\begin{array}{lcr}
 \max_{x \in E} (  Q_x(p) ) =0.82,  & \max_{x \in E} ( \hat{\hat{Q}}_x(p) )=0.42, & err = 5.4\% . 
\end{array}  
\end{equation*}   
As the estimated maximum value ($0.42$) has a large error with respect to the value of the exact solution ($0.82$), we strongly suspect that the quantile function metamodel cannot be directly applied to solve the optimization problem.


In the next section, we will also see that this strategy does not work either on the VME application.
We  then propose an adaptive algorithm which consists in sequentially adding simulation points in order to capture interesting quantile functions to be added in our functional basis.

\subsection{QFEI: an adaptive optimization algorithm }

After the choice of $\chi$, $E$ and the families $\left( Q_x \right)_{x \in \chi}$, $( \hat{Q}_x)_{x \in \chi}$ and $( \hat{\hat{Q}}_x)_{x \in E}$, our new algorithm will propose to perform new interesting (for our specific problem) calls to the VME simulator on $E$ (outside of $\chi$).
With the Gaussian process metamodel, which provides a predictor and its uncertainty bounds, this is a classical approach used for example in black-box optimization problem (Jones et al. \cite{jonsch98}) and rare event estimation (Bect et al. \cite{becgin12}).
The goal is to provide some algorithms which mix global space exploration and local optimization.

Our algorithm is based on the so-called EGO (Efficient Global Optimization) algorithm (Jones et al. \cite{jonsch98}) which uses the Expected Improvement (EI) criterion to maximize a deterministic simulator.
Our case is different as we want to maximize:
\begin{equation}   
\begin{array}{rccl}
H  \, : \,& E & \rightarrow & \mathbb{R} \\
& x & \mapsto & Q_x(p) \quad \mbox{$p$-quantile of NPV$(x)$} ,
\end{array}  
 \end{equation}
 \emph{ie} the $p$-quantile function, for $p \in ]0,1[$, of the stochastic simulator.
We will then propose a new algorithm called the QFEI (for Quantile Function Expected Improvement) algorithm.

As previously, we use a restricted set $E$ with $\#E=5000$ ($E$ is a random sample in the full set), the initial learning set $\chi \subset E$ with $\# \chi=200$ (initial design of experiment), $\left( Q_x \right)_{x \in \chi}$, $( \hat{Q}_x)_{x \in \chi}$ and $( \hat{\hat{Q}}_x)_{x \in E}$. 
We denote the current learning set by $D$ (the initial learning set increased with additional points coming from QFEI).
The Gaussianity on the components of $\psi$ is needed for the EGO algorithm,  that is why we do not perform the logarithmic transformation presented in  Section \ref{sec:metamodelcoef}.
In our case, it has not implied negative consequences.

We apply the Gaussian process metamodeling on the $k$ independent components $\psi_1,...,\psi_k$ (see Equation \eqref{meta_psi}).
As $\hat{Q}_x(p) = \sum_{j=1}^k \psi_j(x) R_j(p) $, we have
\begin{equation}
\hat{Q}_x(p) \sim \mathcal{N} \left( \hat{\hat{Q}}_x(p), \sum_{j=1}^k R_j(p)^2 MSE_j(x)  \right)  \;,\; \forall x \in E. 
\end{equation} 
Then $\hat{Q}_{x}(p)$ is a realization of the underlying Gaussian process $U_{x}=\sum_{j=1}^k \psi_j (x) R_j(p)$ with
\begin{equation} 
\begin{array}{lll}
U_D & := \left( U_x \right) _{x \in D} &,\\
\hat{U}_{x} & := \ee[U_{x} \mid U_D], &\forall x \in E ,\\
\sigma_{U|D}^2(x) & := \mbox{Var}[U_{x} \mid U_D], &\forall x \in E .
\end{array} \end{equation} 
The conditional mean and variance of $U_x$ are directly obtained from the $k$ Gaussian process metamodels of the $\psi$ coefficients \eqref{meta_psi}.

At present, we propose to use the following improvement random function:
\begin{equation}   
\begin{array}{rccl}
I  \, : \,& E & \rightarrow & \mathbb{R} \\
& x & \mapsto & \left(U_x-\max\left(U_D \right)\right)^+ , 
\end{array}  
 \end{equation}   
where $\left( x \mapsto \left(x\right)^+ = x. \mathbf{1}_{x > 0} \right)$ is the positive part function.
In our adaptive design, finding a new point consists in solving:
\begin{equation}  x_{new} := \argmax_{x \in E} \ee[I(x)] .  \end{equation} 
Added points are those which have more chance to improve the current optimum. 
The expectation of the improvement function writes (the simple proof is given in Browne \cite{bro14}):
\begin{equation}  
\quad \ee[I(x)]=\sigma_{U|D}(x) \left( u(x) \phi(u(x)) + \varphi(u(x)) \right) \;,\; \forall x \in E ,\quad \mbox{with} \quad u(x)= \frac{\hat{U}_x-\max(U_D)}{\sigma_{U|D}(x)}   
\end{equation} 
where $\varphi$ and $\phi$ correspond respectively to the density and distribution functions of the reduced centered Gaussian law. 

In practice, several iterations of this algorithm are performed, allowing to complete the experimental design $D$.
At each iteration, a new projection functional basis is computed and the $k$ Gaussian process metamodels are re-estimated.
The stopping criterion of the QFEI algorithm can be a maximal number of iterations or a stabilization criterion on the obtained solutions.
No guarantee on convergence of the algorithm can be given.
In conclusion, this algorithm provides the following estimation of the optimal point $x^*$:
\begin{equation}\label{eq:QFEI_sol}
 \hat{x}^* := \argmax_{x \in D}(U_D) .
\end{equation} 

\subsection{Validation of the QFEI algorithm on the toy function}

We get back to the toy-function $G$ introduced in (\ref{eq:defG}).
The goal is to determine $x^* \in E$ such that:
$$
x^* = \underset{x \in E}{\argmax} \; Q_x(p)
$$
with $p=0.4$.
In other words, we try to find the input $x \in E$ whose output distribution has the highest 40\%-quantile.
In this very case, we have:
$$
x^* = \left( 1, 0,1 , 0.5 \right) \quad \mbox{with} \quad Q_{x^*}(p)= 0.884.
$$
We have also computed
\begin{equation*} 
 \frac{1}{\# E} \sum_{x \in E} Q_x(p)=-0.277, \quad \mbox{Var} \left[ \left(Q_x(p)\right)_{x \in E} \right] = 0.071 . 
\end{equation*} 
Let us first remember that we set an efficient metamodel $\hat{\hat{Q}}$ for $Q$ in the  section \ref{sec:applGP}.
Indeed, we had $err3=1.42 \%$.

As before, we test the natural way to get an estimation for $x^*$, by determining
$$
\tilde{x}^* = \underset{x \in E}{\argmax} \hat{\hat{Q}}_x(p).
$$
Unfortunately, still in our previous example, we get
$$
\tilde{x}^* = \left( 0.9, 0,2, 0.8 \right)
$$
which is far from being satisfying.
Besides, when we compute the real output distribution for $\tilde{x}^*$, we have
$$
Q_{\tilde{x}^*}(p) = 0.739 .
$$
Therefore only relying on $\hat{\hat{Q}}$ to estimate $x^*$ would lead to an important error.
This is due to the high smoothness of the function $x \longrightarrow Q_x(p)$: even a small error in the estimator $Q_{\cdot}(p)$ completely upsets the order of the family $\left( Q_x(p) \right)_{x \in E}$.

At present, we use the QFEI algorithm (Eq. (\ref{eq:QFEI_sol})) in order to estimate $x^*$ by $\hat{x}^* $, with $20$ iterations.
At the end of the experiments, the design $D$ is the learning set $\chi$ to which we have consecutively added $20$ points of $E$. 

With QFEI, we finally get
$$
\hat{x}^* = \left( 1, 0,1 , 0.2 \right) \quad \mbox{with} \quad Q_{\hat{x}^*}(p)= 0.878.
$$
It is important to mention that $\hat{x}^*$ has the second highest output 40\%-quantile. 

Overall we tested the whole procedure $30$ times with, in each case, a new learning set $\chi$.
In the $30$ trials, we made sure that the maximizer $x^*$ was not already in $\chi$.
In $22$ cases, we obtain $\hat{x}^*=x^*$, which is the best result that we could expect.
For the remaining $8$ times, we obtain $\hat{x}^* = \left( 1, 0,1 , 0.2 \right)$.
We can conclude that the QFEI algorithm is really efficient for this optimization problem.

\section{Application to the VME case study}

\begin{figure}[!ht]
\begin{center}
\includegraphics[width=8cm,height=6cm]{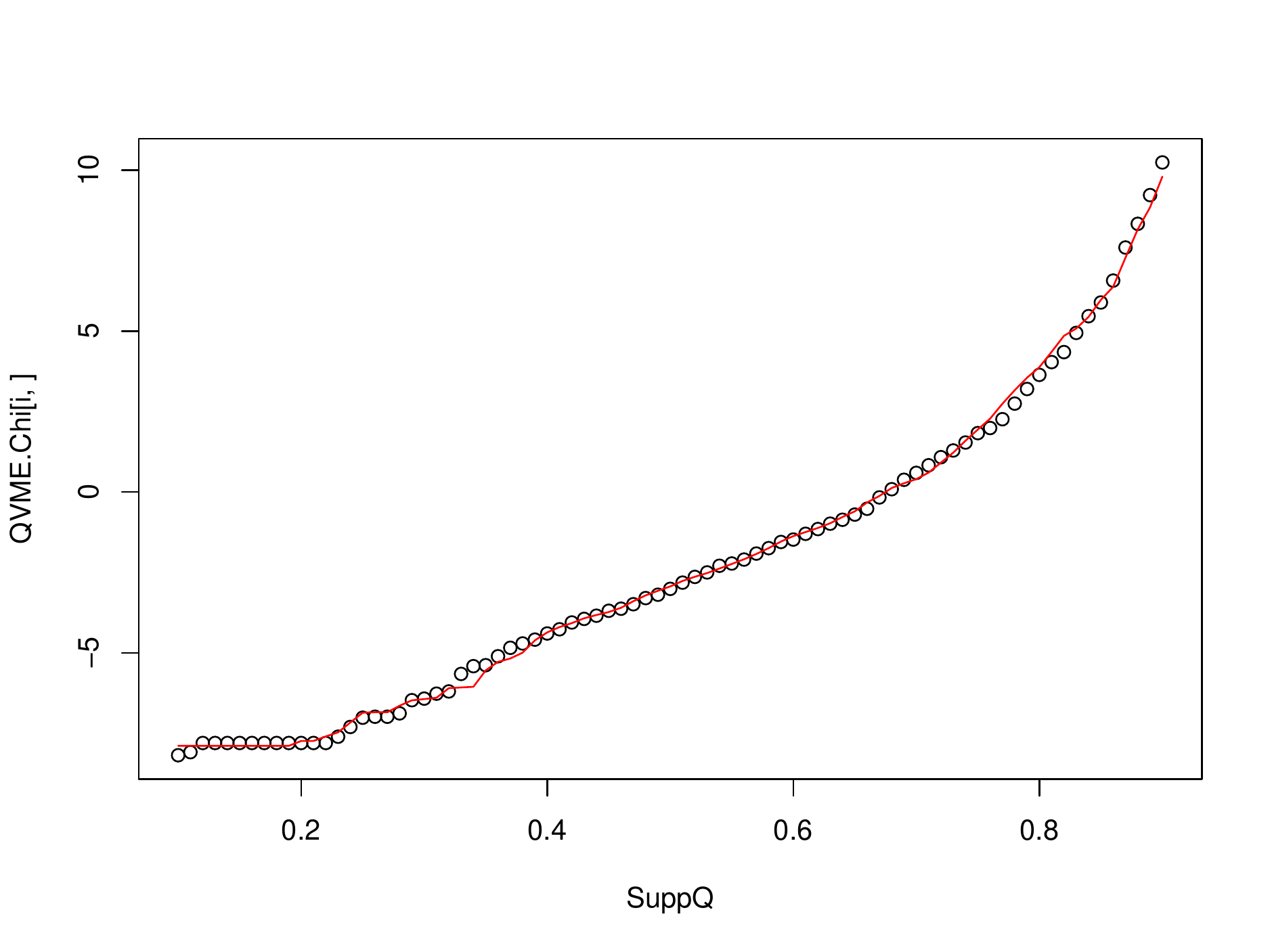}
\caption{For one point $x \in \chi$, $\hat{Q}_x$ (red line) and $Q_x$ (black points).} \label{plot6}
\end{center}
\end{figure}

We return to our VME application (cf Section \ref{sec:VME}).
For the projection of $Q_x$ by the Modified Magic Points (MMP) algorithm, the choice $k=5$ has shown sufficient approximation capabilities.
For one example of quantile function output, a small relative $L^2$-error ($0.2\%$) between the observed quantile function and the projected quantile function is obtained.
Figure \ref{plot6} confirms also the relevance of the MMP method.

\begin{figure}[!ht]
\begin{center}
\includegraphics[width=8cm,height=6cm]{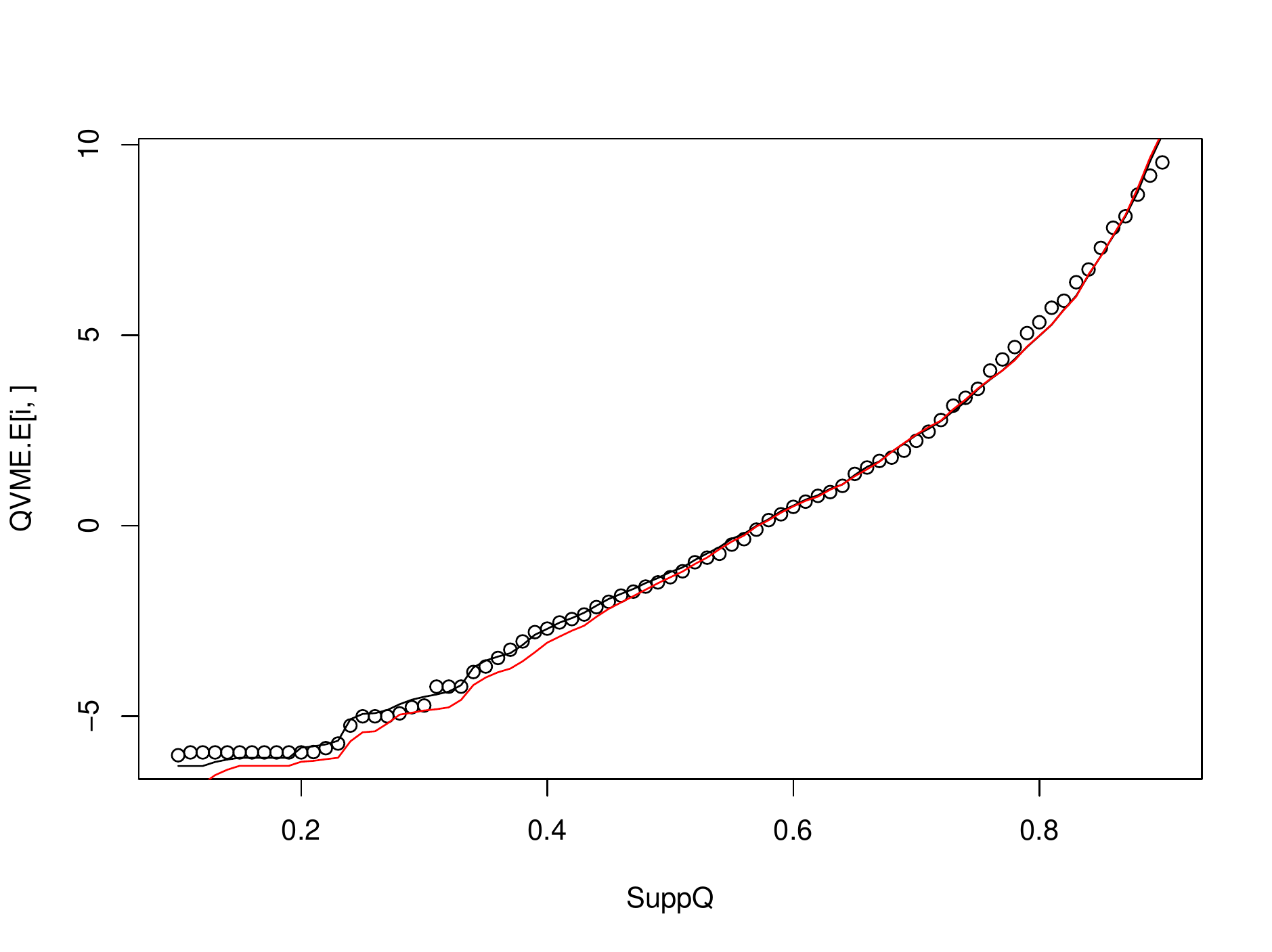}
\caption{For one point $x \in \chi$, $\hat{\hat{Q}}_x$ (red line) and $Q_x$ (black points).}\label{plot7}
\end{center}
\end{figure}

We build the Gaussian process metamodel on the set $E$ (with the choice $k=5$).
For one example of quantile function output, a small relative $L^2$-error ($2.8\%$) between the observed quantile function and the emulated quantile function is obtained.
Figure \ref{plot7} confirms also the relevance of the metamodeling method.

As for the toy function, our optimization exercise is to determine $x^* \in E$ such that
$$
x^* = \underset{x \in E}{\argmax} \; Q_x(p)
$$
with $p=0.5$.
We first try to directly apply an optimization algorithm on the previously obtained metamodel.
As an example, for $p=0.5$ (median estimation), we find:
\begin{equation}   
\begin{array}{lcr}
 \max_{x \in E} (  Q_x(p) ) =0.82,  & \max_{x \in E} ( \hat{\hat{Q}}_x(p) )=0.42, & err = 5.4\% . 
\end{array}  
\end{equation}   
If we define $y= \argmax_{x \in E} [\hat{\hat{Q}}_x(p)]$ the best point from the metamodel, we obtain $Q_y(p)=0.29$ while $\max_{x\in \chi}Q_x(p)=0.35$. 
The exploration of $E$ by our metamodel does not bring any information. 
We have observed the same result by repeating the experiments $100$ times (changing the initial design each time).
It means that the punctual errors on the quantile function metamodel are too large for this optimization algorithm.
In fact, the basis functions $R_1,...,R_5$ that the MMP algorithm has chosen on $\chi$ are not able to represent the extreme parts of the quantile functions of $E$.
As a conclusion of this test, the quantile function metamodel cannot be directly applied to solve the optimization problem. 

We now apply the QFEI algorithm.
In our application case, we have performed all the simulations in order to know $\left( Q_x \right) _{x \in E}$, therefore the solution $x^*$.
Our first objective is to test our proposed algorithm for $p=0.4$ which has the following solution:
\begin{equation}  \left\lbrace
\begin{array}{ll}
x^* & = \left( 41,47,48,45,18 \right) \\
Q_{x^*}(p) & = -1.72 .
\end{array}
\right. \end{equation} 
We have also computed
\begin{equation}  \frac{1}{\# E} \sum_{x \in E} Q_x(p)=-3.15, \quad \mbox{Var} \left( \left(Q_x(p)\right)_{x \in E} \right)=0.59 . \end{equation} 
We start with $D := \chi$ and we obtain
\begin{equation} \max_{x \in \chi} \left( Q_x \right) = -1.95 . \end{equation} 
After $50$ iterations of the QFEI algorithm, we obtain:
\begin{equation}  \left\lbrace
\begin{array}{ll}
\hat{x}^* & = \left( 41,47,45,46,19 \right) \\
Q_{\hat{x}^*}(p) & = -1.74 .
\end{array}
\right. \end{equation} 
We observe that $\hat{x}^*$ and $Q_{\hat{x}^*}(p)$ are close to $x^*$ and  $\simeq Q_{x^*}(p)$.
This is a first confirmation of the relevance of our method. 
With respect to the initial design solution, the QFEI has allowed to obtain a strong improvement of the proposed solution.
$50$ repetitions of this experiment (changing the initial design) has also proved the robustness of QFEI.
The obtained solution is always one of the five best points on $E$.

QFEI algorithm seems promising but a lot of tests remain to perform and will be pursued in future works: changing $p$ (in particular testing extremal cases), increasing the size of $E$, increasing the dimension $d$ of the inputs, \ldots

\section{Conclusion}

In this paper, we have proposed to build a metamodel of a stochastic simulator using the following key points: 
\begin{enumerate}
\item Emulation of the quantile function which proves better efficiency for our problem than the emulation of the probability density function; 
\item Decomposition of the quantile function in a sum of the quantile functions coming from the learning sample outputs; 
\item Selection of the most representative quantile functions of this decomposition using an adaptive choice algorithm (called the MMP algorithm) in order to have a small number of terms in the decomposition; 
\item Emulation of each coefficient of this decomposition by a Gaussian process metamodel, by taking into account constraints ensuring that a quantile function is built.
\end{enumerate}

The metamodel is then used to treat a simple optimization strategy maintenance problem using a stochastic simulator (VME), in order to optimize an output (NPV) quantile. 
Using the Gaussian process metamodel framework and extending the EI criterion to quantile function, the adaptive QFEI algorithm has been proposed. 
In our example, it allows to obtain an ``optimal'' solution using a small number of VME simulator runs.

This work is just a first attempt and needs to be continued in several directions:
\begin{itemize}
\item Consideration of a variable $N_{\mbox{\scriptsize MC}}$ whose decrease could help to fight against the computational cost of the stochastic simulator,
\item Improvement of the initial learning sample choice by replacing the random sample by a space filling design (Fang et al. \cite{fanli06}),
\item Algorithmic improvements to counter the cost of the metamodel evaluations and to increase the size of the study set $E$, 
\item Multi-objective optimization (several quantiles to be optimized) in order to  take advantage of our powerful quantile function emulator,
\item Including the estimation error induced in practice by $\hat{h}$, $\hat{\beta}$ and $\hat{\theta}$ to define a more precise version of the QFEI algorithm,
\item Application to more complex real cases,
\item Consideration of a robust optimization problem where environmental input variables of the simulator has not to be optimized but just create an additional uncertainty on the output.
\end{itemize}



\end{document}